\documentclass[pdflatex,sn-nature]{sn-jnl}% Nature Portfolio style (numbered references)

\usepackage{graphicx}%
\usepackage{multirow}%
\usepackage{amsmath,amssymb,amsfonts}%
\usepackage{booktabs}%

\begin{document}

\title[Lecture board readability and perceived teaching clarity]{Lecture Board Readability as an AI-Measurable Predictor of Perceived Teaching Clarity}

\author[1]{\fnm{Nachshon} \sur{Hirschenberger}}

\author[1]{\fnm{Roni} \sur{Segal}}

\author*[1]{\fnm{Yossi} \sur{Ben Zion}}\email{benzioy@biu.ac.il}

\affil*[1]{\orgdiv{Department of Physics}, \orgname{Bar-Ilan University}, \orgaddress{\city{Ramat Gan}, \postcode{52900}, \country{Israel}}}

\abstract{Lecture boards remain central to university teaching in physics and mathematics, yet the visual quality of what instructors write has rarely been measured systematically. We examined whether a vision-language model can score board writing in a way that relates to how students perceive instruction. We analysed 2,645 board images from 56 university physics and applied mathematics courses taught by 18 lecturers, scoring each image 50 times with Claude 3.5 Sonnet on readability, organization and visual elements. Course-level readability was associated with student-rated clarity (\textit{r}~=~.43) and organization (\textit{r}~=~.46), and remained the only significant predictor of the overall survey score after controlling for class size, mean course grade and study year. These associations were robust to clustering by lecturer and remained stable when only one or two images per lecture were used. In a subsample of 100 images rated by four human raters, agreement between AI and human scores was comparable to agreement among the raters themselves for visual elements, lower for readability and weakest for organization. Student evaluations served as a benchmark of perceived clarity rather than of teaching quality as a whole. Automated analysis of board images may offer a low-cost, scalable source of formative feedback in STEM teaching.}


\maketitle

    \section{Introduction}
    \label{sec:intro}     

    Can a single picture of a classroom lecture board reveal something about instructional clarity that a student survey can capture only once a semester, if at all?

The rapid digitization of higher education has turned boards, slides, and lecture recordings into abundant sources of data. With recent advances in artificial intelligence (AI), especially vision-language models that integrate image interpretation with large language models (LLMs), these instructional artifacts can now be analyzed directly, enabling systematic and scalable measurement of properties that were previously accessible only through individual impression.

Student evaluations of teaching (SETs) remain the most established institutional indicator of how students experience instruction, although questions about their validity remain unresolved \citep{spooren2013validity}; they serve throughout this study as the reference measure against which the automated scores are compared. For the purposes of this study, their main constraint is practical: administering a survey requires institutional coordination, student time, and a response rate high enough to represent the enrolled cohort \citep{nulty2008adequacy}, which in practice restricts collection to a single administration at the end of the course.
Instructors therefore receive one retrospective snapshot per course, with no mechanism for observing their practice while it is still under way. Automated analysis of instructional artifacts is not subject to this constraint: once an evaluation pipeline is established, the same assessment can be applied to every lecture at negligible marginal cost. The two approaches are thus complementary rather than competing, and we use SETs here as a benchmark for student-perceived clarity rather than as an absolute measure of teaching effectiveness.

This study applies such analysis to the lecture board, treating it as an instructional artifact whose visual properties can be measured automatically. Focusing on properties such as legibility, layout, and visual clarity, we examine whether board writing constitutes a measurable visual signal that aligns with students' perceptions of teaching clarity and organization. To further validate the AI-based scoring procedure, we additionally compare it against independent human ratings. A supplementary, exploratory extension examining the integration of board-based visual metrics with lecture-transcript data is presented in Supplementary Note 1.

\subsection{Board Writing in the Exact Sciences}
\label{sec:board_writing}

Ball~\cite[p.~34]{ball2017power} argues that ``blackboard and chalk -- like paper and ink -- are a combination that modern technologies can't improve or displace,'' noting their central position in theoretical physics despite sophisticated digital alternatives. Empirical support comes from a survey of mathematics lecturers at a South African university by Billman et al.~\cite{billman2018does}, in which half of the respondents agreed that the chalkboard is still the best way to teach mathematics. These views were reported even though chalkboard use in the department had declined significantly over the preceding decade, as the teaching of large groups increasingly required technology.

The deliberate pace of board writing may support the sequential, step-by-step nature of mathematical proofs and derivations. Findings by Wecker~\cite{wecker2012slide} on a ``speech suppression'' effect provide relevant insight: in an experiment with university students, information presented only orally was retained less well when the presentation was accompanied by conventional slides, an effect attributed not to cognitive overload but to learners allocating their attention to the slides, and avoided when the slides were concise. Since mathematical instruction relies heavily on oral explanations accompanying the written development of formulas and solutions, board writing, in which oral explanation and writing develop together rather than being displayed in advance, may reduce this problem, although this possibility has not been tested directly. A meta-analysis of 48 studies by Baker et al.~\cite{baker2018does} found no overall difference in learning between PowerPoint-based and traditional instruction, and no advantage for college students.

Ball~\cite{ball2017power} further emphasizes the collaborative dimension of blackboard use, describing the blackboard as ``a democratic space, where ideas can be easily shared'' (p.~35), and notes that newer writing surfaces introduced at research institutions were often abandoned in favour of traditional blackboards, which he describes as ``an optimized technology'' (p.~35). This persistence raises an important question: if lecture boards remain so central to teaching in the exact sciences, why have they received so little attention as tools for pedagogical assessment?

Board writing has nonetheless been studied as an instructional practice. In an international study of undergraduate mathematics teaching in seven countries, Artemeva and Fox~\cite{artemeva2011writing} identified ``chalk talk,'' in which the lecturer develops a mathematical argument in writing on the board while explaining it aloud, as the central pedagogical genre of the mathematics lecture, and found that this genre appeared to override local differences between instructional contexts. Analyzing video recordings of lectures in formal logic, Greiffenhagen~\cite{greiffenhagen2014materiality} showed that presenting mathematics at the board relies on specific inscription practices, such as the use of lines and boxes and the designation of particular board regions for specific mathematical purposes, and argued that writing is integral to mathematical thinking itself. The board also shapes what students take away from a lecture: in proof-based mathematics courses, students' notes typically included what the lecturer wrote on the board but not what was only said aloud \citep{fukawa2017informal}. As with slides, students thus privilege written over spoken content, which makes the quality of what is written on the board consequential for what they take away.

    \subsection{Traditional Evaluation Methods}

    Efforts to evaluate teaching quality have extended beyond student surveys, drawing on a wide range of traditional methods. Martinez et al.~\cite{martinez2016classroom} examined 16 classroom observation systems across six countries and found that, although the systems were consistent in their stated purposes, they varied in how they operationalized good teaching, in the degree of standardization of the observation process, and in their emphasis on validation. Rater consistency poses a further challenge: in a study of principals' classroom observations at the conclusion of training, Bergin et al.~\cite{bergin2017teacher} found that 1,324 trained evaluators, each rating four ten-minute classroom videos, achieved high overall accuracy but varied considerably in individual accuracy, indicating that rater accuracy cannot be assumed after initial training and depends on continued training and practice. Rater-level variation of this kind also emerged among the human raters in the present study (Section~\ref{sec:human_validation}).

Outcome-based approaches face similar limitations: value-added models that link student test-score gains to teacher performance rely on assumptions that rarely hold in practice and produce estimates that are unstable across years \citep{darling2015can}, and research linking teaching to outcomes indicates that effective teaching is multidimensional \citep{kyriakides2013matters}.

Although much of this research concerns school teaching, the constraints it identifies, namely the cost and limited standardization of observation, variation among raters, and the instability of outcome-based estimates, are also relevant to university instruction. These challenges underscore the need for evaluation approaches that are actionable, scalable, and context-sensitive, providing the backdrop for exploring digital and AI-based alternatives.

    \subsection{Digital and AI-Based Approaches}

    The digitization of higher education introduced learning analytics, which draws on data such as learning management system (LMS) traces and questionnaires to describe students' engagement and experience. In a systematic review of 34 studies published between 2015 and 2024 on the use of learning analytics to evaluate course effectiveness in higher education, Wong et al.~\cite{wong2025role} found that descriptive analysis was the most common technique, although machine learning and other AI methods were becoming more prominent, and identified limitations including small samples, biases associated with survey-based methods, and difficulty in generalizing findings across disciplines. Such data capture what students do and report, but provide limited access to the instructional processes behind these outcomes. Teaching analytics extends this work to the teaching act itself, using data and visual analytics to support instructors' reflection on their practice \citep{ndukwe2020teaching}, and video-based observation has been reviewed as a tool for monitoring and evaluating the impact of development projects \citep{borg2021video}.

Artificial intelligence has extended these possibilities, but its use in higher education has focused mainly on students. In a systematic review of 138 studies of AI in higher education published between 2016 and 2022, Crompton and Burke~\cite{crompton2023artificial} found that assessment and evaluation was the most common application, yet 72\% of the studies focused on students and only 17\% on instructors. Studies that do target teaching have typically combined several behavioral and administrative data streams. Guo et al.~\cite{guo2021ai} developed a classroom evaluation model for a smart-classroom environment in China that applied computer vision and speech recognition to recordings of teachers and students, combining statistical weighting with ensemble learning to evaluate indicators such as students' concentration and participation and teachers' style and media use. 
In school settings, Zhou et al.~\cite{zhou2025teaching} developed a teaching quality evaluation framework for blended classrooms that integrates multi-source data from physical and digital learning environments in a multimodal knowledge graph, uses graph attention networks for evaluation, and generates personalized teaching recommendations; the framework was applied to more than 70,000 primary and secondary school students and nearly 8,000 teachers in western China. Closest to the present study, Ben Zion et al.~\cite{zion2025ai} applied language models to transcripts of 1,222 hours of physics lectures and obtained consistent measures of clarity and structure that correlated with student surveys, while noting that transcript-based analysis is domain-specific and excludes non-verbal cues. Computer-vision research on lecture boards has focused on extracting and indexing handwritten content from whiteboard and chalkboard lecture videos for summarization and retrieval \citep{davila2021fcn}, rather than on evaluating how that content is presented. None of these approaches assesses the visual quality of what instructors write during the lecture.

The present study uses a vision-language model as an evaluator, an approach known in the machine-learning literature as LLM-as-a-judge. Zheng et al.~\cite{zheng2023judging} found that strong LLM judges such as GPT-4 matched human preferences for chat-assistant responses with over 80\% agreement, the same level of agreement observed between humans, while also documenting biases such as sensitivity to the position and verbosity of the responses being compared. For multimodal models, Chen et al.~\cite{chen2024mllm} found that model judgments were close to human judgments in pairwise comparisons but diverged substantially from them in direct scoring, and reported biases, hallucinations, and inconsistent judgments even in advanced models. In education, Wang and Demszky~\cite{wang2023chatgpt} found that ChatGPT's zero-shot scores of elementary mathematics classroom transcripts on established observation instruments correlated weakly with those of expert human raters. Because the outputs of such models vary across runs, evaluation frameworks have aggregated repeated samples; Liu et al.~\cite{liu2023geval}, for example, sampled each GPT-4 rating 20 times to estimate the probabilities of the possible scores. These considerations are reflected in two features of the present design: averaging 50 independent runs per image, and validating the resulting scores against independent human raters.
    
    \subsection{Board Writing as a Source of Evidence on Perceived Clarity}

Research on teaching evaluation has drawn on multiple sources of evidence, from linguistic features in lecture transcripts \citep{zion2025ai} and nonverbal cues in video recordings \citep{ambady1993half, guo2021ai} to engagement metrics derived from learning management systems \citep{wong2025role}. Board writing, which remains central to instruction in the exact sciences (Section~\ref{sec:board_writing}), has received far less attention as a source of such evidence. Studies of board writing have described it as a pedagogical genre \citep{artemeva2011writing}, as a spatially organized form of inscription \citep{greiffenhagen2014materiality}, and as the content that students copy into their notes \citep{fukawa2017informal}. To our knowledge, however, no study has systematically measured the visual quality of board writing or examined how it relates to students' perceptions of the clarity and organization of instruction. Where lectures are recorded, the board is captured as a matter of routine, which makes it an untapped opportunity for systematic, AI-based measurement of how instructional content is visually presented, and a way of connecting a traditional teaching practice with current analytical tools.

There are also theoretical grounds to expect that visual properties of board writing, such as its legibility, could bear on how students judge the clarity of instruction. Teacher clarity has been studied extensively in relation to student learning: in two meta-analyses, Titsworth et al.~\cite{titsworth2015two} found that clarity had a moderate effect on students' affective and cognitive learning, accounting for approximately 13\% of the variance in learning outcomes in the first analysis, and recommended a shift from high-inference perceptual measures of clarity toward low-inference behavioral measures. Legibility, in turn, has been shown to affect evaluations of handwritten material through the fluency with which it is processed \citep{greifeneder2010writing}. Board legibility is therefore a plausible, observable correlate of perceived instructional clarity, and measuring it directly from board images is consistent with the call for lower-inference measures of clarity. Possible mechanisms for such a link are considered in the Discussion.

A further consideration concerns how much evidence such a measure requires. In research on thin slices of behavior, Ambady and Rosenthal~\cite{ambady1993half} found that strangers' judgments of college teachers' nonverbal behavior, based on silent video clips totaling 30 seconds per teacher, correlated strongly with end-of-semester student evaluations ($r = .76$ for 13 teachers), and that predictive accuracy did not differ significantly when the clips were shortened to 15 or 6 seconds. These findings have two implications for the present study. First, a small sample of instructional behavior may carry a stable signal of how a course is experienced, which motivates testing whether a few board images per lecture are sufficient. Second, because the clips were silent, the findings indicate that student evaluations partly reflect impressions formed independently of instructional content, a consideration relevant to the present measure, which likewise assesses how content is presented rather than the content itself.

\begin{figure}[ht]
        \centering
        \includegraphics[width=0.85\textwidth]{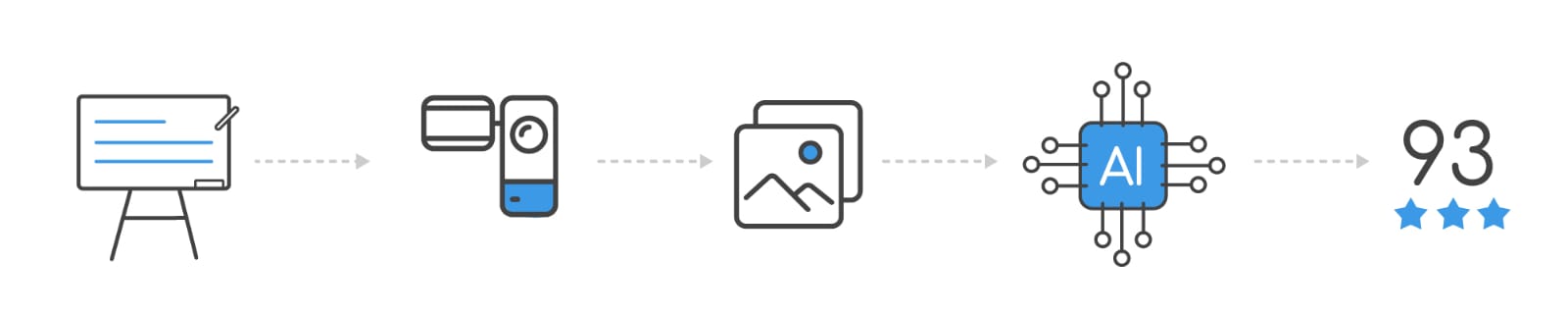}
        \caption{An overview of the AI-based board evaluation process. The lecture board is visually captured, converted into a sequence of images, analyzed by an AI model, and assigned structured visual scores.}
        \label{fig:ai_process}
    \end{figure}

In this study, we analyze 2,645 lecture board images from 56 university physics and applied mathematics courses delivered between 2017 and 2023. The images were sampled at fixed time points during each lecture, and each image was scored 50 times by a vision-language model on three visual dimensions using a structured prompt. The resulting scores were compared with institutional student evaluations from courses with adequate survey response rates, using correlation analyses, control for course-level covariates, and robustness tests based on random and thin-slice sampling. The study aims to determine the extent to which AI-based visual assessments of board writing align with student-perceived clarity and organization of teaching, and whether such an approach could complement existing evaluation methods in higher education. The evaluation process is illustrated in Figure~\ref{fig:ai_process}.
    
    These considerations lead to two research questions, concerning the convergence of AI-based board assessments with student perceptions and the stability of this convergence under thin-slice sampling:
    \begin{enumerate}
        \item To what extent do AI-based visual assessments of lecture board images correspond to students' perceptions of teaching clarity and organization, as reflected in institutional surveys?
        \item To what extent do these AI-based board assessments remain stable predictors of student perceptions when derived from thin-slice samples, that is, from a reduced number of images per lecture?
    \end{enumerate}

    \section{Results}

    Each of the 56 courses was characterized by the mean AI-based scores of its board images on three visual dimensions, referred to below as \textit{Readability} (readability and clarity), \textit{Order} (organization and layout) and \textit{Visuals} (visual elements and illustration). These scores were compared with six measures from institutional student evaluations of teaching: \textit{Clarity}, \textit{Interest}, \textit{Organization}, \textit{Questions}, \textit{Satisfaction} and the composite \textit{Average} (see Methods, Section~\ref{sec:Methodology}).

    \subsection{Correlations between Survey and AI-based Measures}
    
    Table \ref{tab:ai_survey} presents the correlations between the survey-based indicators of student-perceived teaching and the three AI-derived board measures. As expected, the survey dimensions were strongly interrelated: \textit{Clarity}, \textit{Organization}, \textit{Interest}, \textit{Questions}, and \textit{Satisfaction} all showed high positive correlations with one another. This pattern indicates a high degree of internal coherence among the survey responses.
    
    Turning to the AI-based measures, \textit{Readability} exhibited the most consistent associations with the survey indicators, showing significant correlations with \textit{Clarity} ($r=.434, p<.01$), \textit{Organization} ($r=.455, p<.01$), \textit{Questions} ($r=.316, p<.05$), \textit{Satisfaction} ($r=.369, p<.01$), and the composite \textit{Average} score ($r=.422, p<.01$). In contrast, the \textit{Order} and \textit{Visuals} measures showed weaker or non-significant relationships with most of the survey dimensions, although \textit{Visuals} was moderately correlated with both \textit{Readability} ($r=.430, p<.01$) and \textit{Order} ($r=.549, p<.01$). The associations of \textit{Readability} with \textit{Clarity}, \textit{Organization} and \textit{Average} remained significant after Bonferroni correction for the 18 AI--survey correlations ($\alpha = .0028$), whereas those with \textit{Satisfaction} and \textit{Questions}, and the association between \textit{Visuals} and \textit{Interest}, did not.
    
    Interestingly, although \textit{Interest} and \textit{Clarity} were strongly correlated within the survey measures ($r=.632, p<.01$), \textit{Interest} did not show a significant association with the AI-based \textit{Readability} score ($r=.226, p=.095$). However, a modest correlation, significant before correction, was observed between \textit{Visuals} and \textit{Interest} ($r=.278, p<.05$).
    
    Overall, the results indicate that while survey responses are highly interdependent, the AI measures capture narrower, visually grounded aspects of the instructional setting, with \textit{Readability} showing the strongest and most consistent alignment with student evaluations. Given this pattern of associations, \textit{Readability} is used as the primary AI-derived indicator in the subsequent analyses.
    
    Consequently, to ensure a robust analysis, the subsequent examination is restricted to the survey dimensions and the composite score that demonstrated the strongest and most consistent associations with \textit{Readability}. Accordingly, we proceed by analyzing \textit{Clarity}, \textit{Organization}, \textit{Satisfaction}, and the \textit{Average} score (all $p < .01$ before correction), while excluding \textit{Questions} and \textit{Interest} from the further stages of the study; results for \textit{Satisfaction}, which did not survive Bonferroni correction, should be read with this in mind.

    \begin{table}[ht]
    \caption{\label{tab:ai_survey}Correlations between survey-based measures of student-perceived teaching and AI-based board measures ($N=56$ courses).}
    \begin{tabular}{@{}lcccccccc@{}}
    \toprule
     & 1 & 2 & 3 & 4 & 5 & 6 & 7 & 8 \\
    \midrule
    \multicolumn{9}{@{}l}{\textit{Survey}} \\
    1. Clarity       &         &        &        &        &        &        &        &        \\
    2. Interest      & .632**  &        &        &        &        &        &        &        \\
    3. Organization  & .812**  & .439** &        &        &        &        &        &        \\
    4. Questions     & .569**  & .388** & .440** &        &        &        &        &        \\
    5. Satisfaction  & .882**  & .755** & .767** & .608** &        &        &        &        \\
    6. Average       & .897**  & .685** & .853** & .614** & .924** &        &        &        \\
    \midrule
    \multicolumn{9}{@{}l}{\textit{AI}} \\
    7. Readability   & .434**  & .226   & .455** & .316*  & .369** & .422** &        &        \\
    8. Order         & $-$.040 & .041   & $-$.032 & $-$.014 & $-$.062 & $-$.071 & .336*  &        \\
    9. Visuals       & .096    & .278*  & $-$.057 & .218  & .058   & .069   & .430** & .549** \\
    \botrule
    \end{tabular}
    \footnotetext{Pearson correlations; column numbers refer to the numbered rows. $^{*}p<.05$, $^{**}p<.01$ (two-tailed). Survey items are shortened for display.}
    \end{table}

    \subsection{Robustness to Reduced Sampling: One or Two Images per Lesson}
    \label{sec:slicing}
    
\begin{figure}[h!]
        \centering
        % width=\textwidth defines the max width
        % height=0.9\textheight defines the max height
        % keepaspectratio says "fit inside these limits without stretching"
        \includegraphics[width=\textwidth, height=0.9\textheight, keepaspectratio]{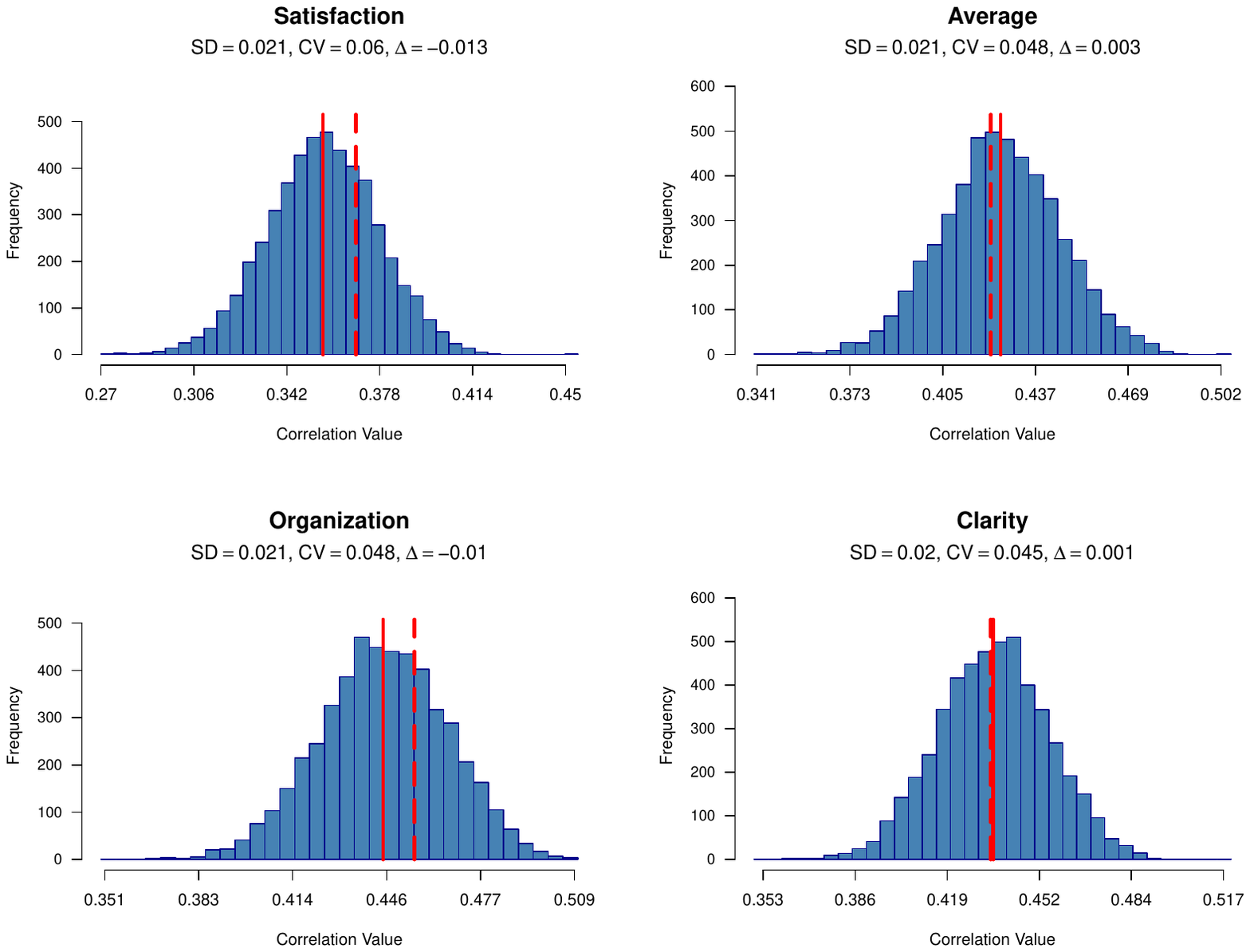}
        \caption{Histograms showing the distribution of correlation values across 5,000 random iterations of selecting two images per lesson. The dashed red line indicates the original correlation value, and the solid red line represents the mean of the simulation.}
        \label{fig:2_pic_corr}
    \end{figure}

    To gain a better understanding of the statistical power of the slicing method and to explore whether the amount of data collected from each course could be further reduced, we conducted the following analysis. From each lesson, we randomly selected two images and calculated the correlation, repeating the random selection and correlation computation 5,000 times. The distribution of correlation values for each survey parameter is shown in Fig.~\ref{fig:2_pic_corr}. As shown, the correlation remained nearly unchanged.
    
\begin{figure}[h!]
        \centering
        % width=\textwidth defines the max width
        % height=0.9\textheight defines the max height
        % keepaspectratio says "fit inside these limits without stretching"
        \includegraphics[width=\textwidth, height=0.9\textheight, keepaspectratio]{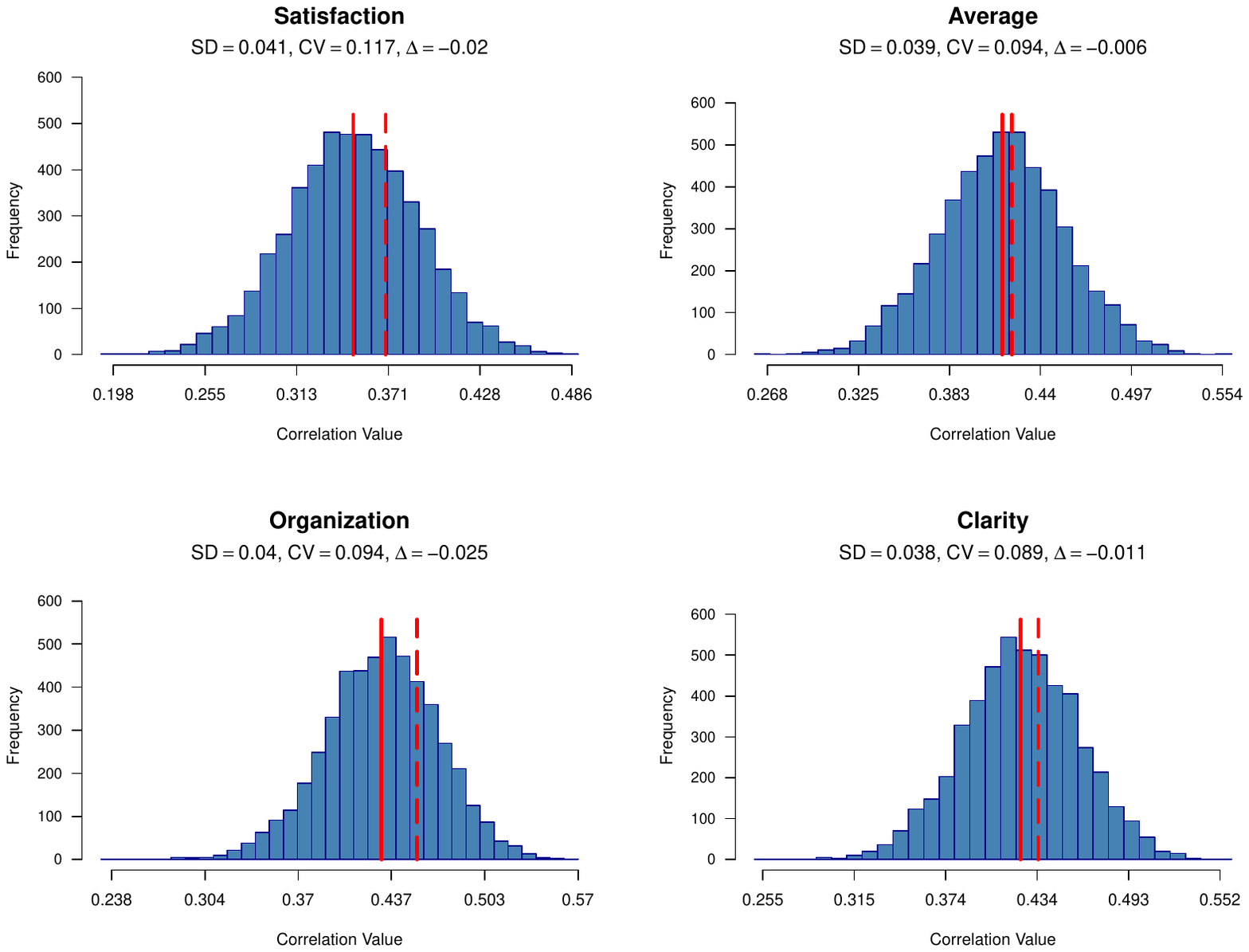}
        \caption{Histograms showing the distribution of correlation values across 5,000 random iterations of selecting a single image per lesson. The dashed red line indicates the original correlation value, and the solid red line represents the mean of the simulated distribution.}
        \label{fig:1_pic_corr}
    \end{figure}

    Subsequently, we examined whether it would be possible to use only a single image from each lesson. In this case as well, the image from each lesson was randomly selected and the whole process was repeated 5,000 times. The corresponding histograms are shown in Fig.~\ref{fig:1_pic_corr}. Overall, these results indicate that selecting a smaller number of images from each lesson preserves the correlation with only minor loss. This provides a stronger justification for the use of the slicing method and supports its statistical stability.
    
    \subsection{Assessing Robustness: Random Sampling Analysis}
    
    To further evaluate the robustness of our findings, we examined whether the total number of images used in the analysis could be reduced even without maintaining the original lesson structure. To do this, we randomly sampled n images from each course, where $n$ ranged from 1 to 78 (the maximum number of images available in a single course; for courses reaching this maximum, all images were included), and calculated the correlation between \textit{readability} and the four survey parameters. This procedure was repeated 5,000 times for every value of n, after which we averaged the resulting correlations.
    
\begin{figure}
        \centering
        \includegraphics[width=1\linewidth]{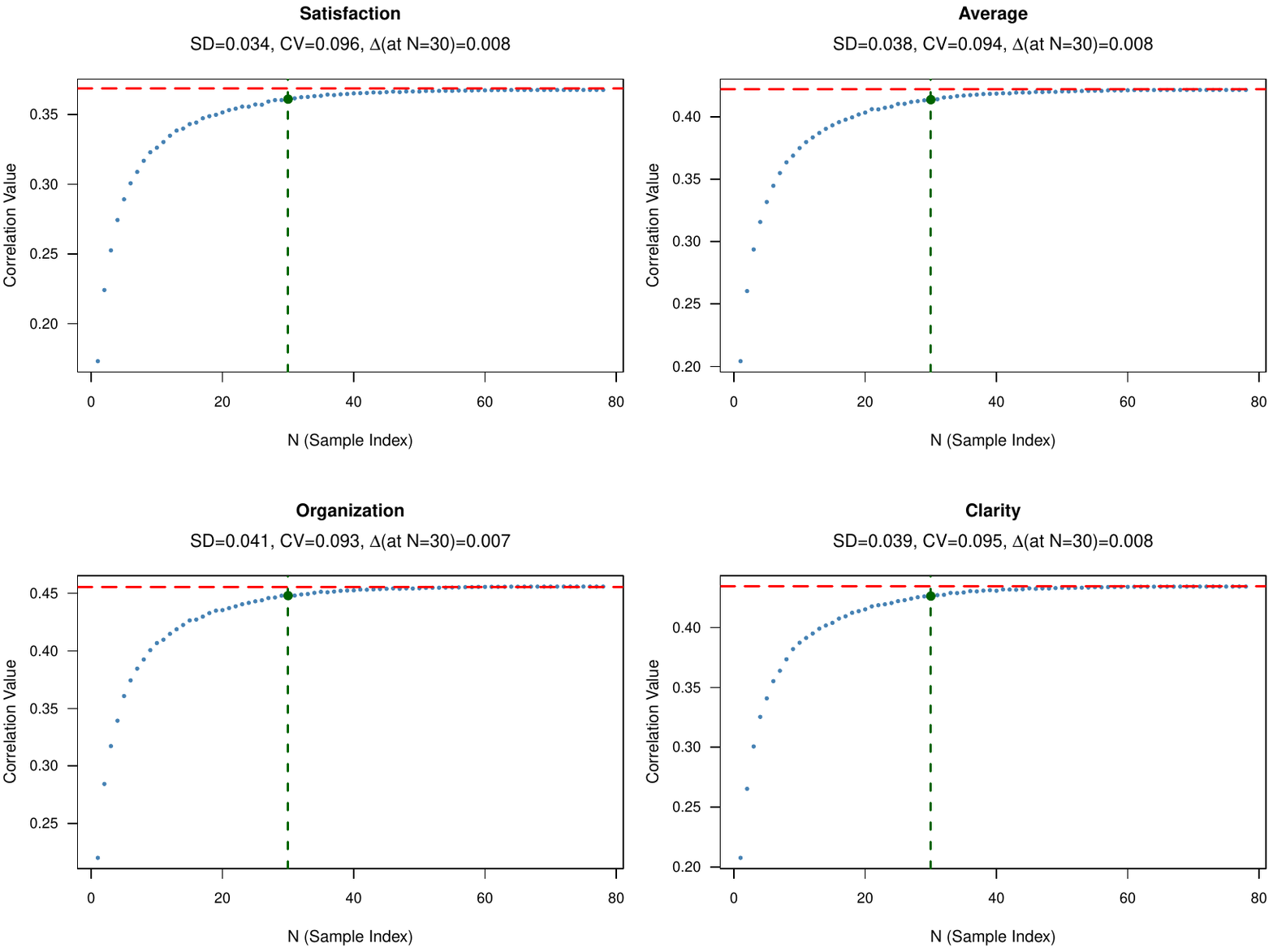}
        \caption{Convergence to correlation as a function of the number of images from each course, with the red line indicating the final correlation. Notably, the method converges to a negligible error margin after only 30 images, compared to an average of 47 available images per course.}
        \label{fig:avg_cor_5000}
    \end{figure}

    Fig.~\ref{fig:avg_cor_5000} presents the correlation as a function of the number of sampled images and provides a clear view of the convergence behavior toward the full data correlation. As shown, the error becomes very small once approximately 25-30 randomly chosen images per course are included. These results demonstrate that a relatively modest number of images is sufficient to recover the essential statistical relationships, thereby reinforcing the validity of the image based characterization of each course.

    \subsection{Controlling for Course-Level Covariates}
    \label{sec:covariates}
    
    The analyses reported so far indicate that AI-assessed \textit{Readability} is associated with student-reported clarity and organization. A remaining question is whether this association simply reflects course characteristics known to shape SET scores independently of instruction. 
    We therefore obtained three course-level covariates from institutional
    records: the mean final grade in the course, the class size (number of students enrolled in the course), and the study year (1--3). Class size and course grade are among the variables most frequently raised as potential confounds in the SET literature \citep{greenwald1997, bedard2008, spooren2013validity}, while study year captures possible differences in evaluation patterns across stages of academic progression. Response rate, a further commonly cited covariate, was not entered into the main model because courses were selected for adequate response rates; adding it as a further covariate did not change the results (\textit{Readability}: $\beta = .42$, $p = .002$).
    
    We estimated a two-step hierarchical linear regression predicting the composite survey score (\textit{Average}). Step 1 included the three covariates; Step 2 added the AI-derived \textit{Readability} score. Results are reported in Table~\ref{tab:covariates}.
    
    \begin{table}[ht]
    \caption{\label{tab:covariates}Hierarchical regression predicting the composite survey score (\textit{Average}), $N = 56$.}
    \begin{tabular}{@{}llrrr@{}}
    \toprule
    Step & Predictor & $B$ & $\beta$ & $p$ \\
    \midrule
    1 & Study year   & $-0.055$ & $-.15$ & .30 \\
      & Class size   & $-0.001$ & $-.11$ & .43 \\
      & Course grade & 0.001    & .03    & .83 \\
      & \multicolumn{4}{l}{$R^2 = .04$, $F(3,52) = 0.64$, $p = .59$} \\
    \midrule
    2 & Study year   & $-0.047$ & $-.13$ & .33 \\
      & Class size   & $-0.001$ & $-.09$ & .50 \\
      & Course grade & 0.000    & $-.00$ & .98 \\
      & Readability  & 0.321    & .41    & .002 \\
      & \multicolumn{4}{l}{$R^2 = .20$, adj.\ $R^2 = .14$, $F(4,51) = 3.25$, $p = .019$} \\
      & \multicolumn{4}{l}{$\Delta R^2 = .17$, $F_{\text{change}}(1,51) = 10.73$, $p = .002$} \\
    \botrule
    \end{tabular}
    \footnotetext{OLS estimates. Class size is the number of students enrolled in the course. Study year is coded 1--3 and treated as continuous.}
    \end{table}

    The baseline model, which included only the traditional course-level predictors, accounted
    for a negligible proportion of the variance and did not reach statistical significance
    ($R^2 = .04$, $F(3,52) = 0.64$, $p = .59$). In particular, mean course grade did not predict
    the survey score, providing no indication of an ``easy course'' effect in this sample.
    Adding \textit{Readability} significantly improved the model ($\Delta R^2 = .17$,
    $F_{\text{change}}(1,51) = 10.73$, $p = .002$), and it remained the only significant predictor ($\beta = .41$, $p = .002$). 
    Its standardized coefficient was virtually identical to its zero-order correlation with the survey score ($r = .42$).

    Taken together, these results indicate that the association between board readability and student-perceived clarity is not attributable to course grade, class size, or academic level. The visual signal carries explanatory power beyond conventional academic factors, capturing
    variation in how students experience instruction that is not recoverable from the administrative course characteristics routinely available to institutions.

    Because several lecturers taught more than one course instance (18 lecturers in total), we repeated these analyses with standard errors clustered by lecturer; the association of \textit{Readability} with student ratings remained significant (all $p < .005$). Averaging scores per lecturer ($N = 18$) yielded correlations of $r = .54$ to $.68$ (all $p < .05$), and the association appeared to be carried mainly by differences between lecturers rather than by variation within the same lecturer, although the small number of lecturers limits the precision of these estimates.

    \subsection{Human Validation of AI-Based Board Assessments}
    \label{sec:human_validation}
    
    To further validate the AI-based scoring procedure, a random sample of 100 lecture board images was drawn from the full image pool used in the primary analysis. Each image was independently rated by four evaluators with a background in the exact sciences, who were not acquainted with the lecturers, were not involved in the research, and were blind to its hypotheses and objectives. Raters scored each image on the same three dimensions used throughout this study -- Readability, Order, and Visuals -- on a 1--10 scale, following the same rating instructions given to the AI model (see Section~\ref{prompt}). Two AI models were included for comparison: Claude 3.5 Sonnet, the primary model used throughout this study, and Gemini 3.5 Flash, included to examine whether the pattern of results generalizes beyond a single AI system. All associations were assessed using Spearman correlations ($\rho$).
    
    Because human judgment of visual board quality is itself not perfectly consistent, a meaningful benchmark for AI performance is not perfect agreement, but the level of agreement observed among human raters themselves. For each rater, we computed the correlation between that rater's scores and the mean score of the other three raters; averaging this value across the four raters yields a ceiling of human agreement for each dimension, against which AI performance can be meaningfully compared (Table~\ref{tab:human_validation}).

    \begin{table}[h!]
    \caption{Agreement between human raters and AI-based scores (Spearman's $\rho$). H1--H4: each rater against the mean of the other three raters (leave-one-out); Ceiling: mean of H1--H4. AI models: each model against the mean of the four raters, and as a percentage of the ceiling.}
    \label{tab:human_validation}
    \begin{tabular}{@{}lcccccrrrr@{}}
    \toprule
     & \multicolumn{5}{c}{Human raters} & \multicolumn{2}{c}{Claude 3.5} & \multicolumn{2}{c}{Gemini 3.5 Flash} \\
     & H1 & H2 & H3 & H4 & Ceiling & $\rho$ & \% & $\rho$ & \% \\
    \midrule
    Readability & .536** & .198* & .592** & .457** & .446 & .287** & 64 & .311** & 70 \\
    Order       & .268** & .436** & .443** & .447** & .399 & .122\phantom{**} & 31 & .447** & 112 \\
    Visuals     & .439** & .448** & .775** & .814** & .619 & .653** & 106 & .761** & 123 \\
    \botrule
    \end{tabular}
    \footnotetext{$^{*}p<.05$, $^{**}p<.01$. \%: percentage of the human ceiling. Human raters and Gemini: $n=100$; Claude: $n=98$ (two images without scores because of API errors).}
    \end{table}
    
    The results reveal that human agreement is far from uniform, and that no single rater is a consistent outlier across dimensions. H2 shows the weakest agreement with the group for Readability, while H1 shows the weakest agreement for Order; on Visuals, agreement is the highest of the three dimensions, although H3 and H4 agree more closely with the group than H1 and H2. This pattern suggests that human observers themselves converge more easily on what constitutes good visual layout than on what constitutes good readability or order, i.e., that these constructs carry a degree of inherent subjective ambiguity independent of the scoring method itself. To our knowledge, this level of disaggregation -- reporting agreement separately for each rater and dimension, rather than a single pooled coefficient -- is uncommon in studies validating AI-based teaching-evaluation tools against human judgment.
    
    The right-hand columns of Table~\ref{tab:human_validation} present the correlation between each AI model's scores and the mean of the four human raters, expressed as a percentage of the human agreement ceiling.

    The pattern of results is consistent with the human agreement findings above. For Claude, alignment with human raters tracks the human agreement ceiling closely: it is weakest for Order, the dimension on which humans themselves agree least, and reaches the human ceiling for Visuals, the dimension of highest human agreement. Gemini shows a broadly similar alignment with human raters overall, reaching or exceeding the human ceiling for both Order and Visuals, and approaching it for Readability. For the dimension on which humans agree most consistently (Visuals), both AI models reach a level of association with human raters comparable to the association raters show with one another, indicating that, at least for this dimension, the AI-based scores fall within the range of human agreement rather than outside it.
    
    \section{Discussion}

A central finding of this study is that the readability of handwritten boards, assessed independently of content, was consistently associated with students' ratings of teaching clarity and organization. One interpretation of this pattern is that judgments of instructional clarity draw not only on pedagogical and linguistic factors but also on the earliest stage of learning from the board: how easily students can visually take in what is written, before they begin to process its meaning. The account below draws on research in visual perception, attention, cognitive load, and processing fluency. Because the present data are correlational, it is offered as a plausible mechanism rather than a tested one.

In physics and applied mathematics, board writing consists of letters, numerals, and mathematical symbols. Before students can follow a derivation, they must recognize the form of each of these elements, a step that is distinct from understanding what they mean. For skilled readers, the recognition of familiar words is largely automatic and places few demands on attention \citep{laberge1974toward}, and visual recognition more generally builds on early stages of processing that extract elementary features such as edges and line orientations \citep{marr2010vision}. Feature Integration Theory \citep{treisman1980feature} distinguishes between the registration of such basic features, which occurs automatically across the visual field, and their integration into coherent objects, which requires focused attention: ``focal attention provides the `glue' which integrates the initially separable features into unitary objects'' \citep[p.~98]{treisman1980feature}. When board writing is clear, the form of each symbol can be recognized with little effort, leaving cognitive resources available for the content. Less legible writing, by contrast, may require more attention and decoding effort before the form of a symbol is established.

This additional effort carries a cost in working memory. Sweller~\cite{sweller1988cognitive} showed that cognitive activities unrelated to the learning goal, such as means--ends problem solving, consume working-memory resources that would otherwise support schema acquisition. Later formulations of cognitive load theory term such processing extraneous load \citep{sweller1998cognitive}. Deciphering degraded handwriting or ambiguous symbols is a clear instance: resources are spent on recovering the form of the material rather than on understanding it.

The consequences of decoding difficulty are not limited to comprehension. Research on processing fluency shows that the subjective ease with which information is processed serves as a cue in judgment: fluently processed stimuli tend to be liked more \citep{reber2004processing} and judged as more likely to be true \citep{reber1999effects}; see Alter and Oppenheimer~\cite{alter2009uniting} for a review. Handwriting is a direct case in point. Greifeneder et al.~\cite{greifeneder2010writing} found that essays written in legible handwriting were evaluated more positively than the same essays in less legible handwriting, and related this effect to the fluency with which the handwriting was processed. Applied to the lecture hall, this suggests that the ease of reading the board may be experienced by students as clarity of the instruction itself.

This account also delimits what the Readability score can be taken to indicate. Fluency effects operate on perceptions, and perceptions do not necessarily track learning: Carpenter et al.~\cite{carpenter2013appearances} found that a fluent instructor delivery increased students' perceived learning without increasing actual learning. The association reported here should therefore be read as a link between board readability and student-perceived clarity, the construct that SETs capture, rather than as evidence that readable boards improve learning outcomes. The association may also partly reflect other instructional habits that co-occur with careful board writing, such as thorough lecture preparation or a measured pace of presentation, which could themselves contribute to students' ratings. The covariate analysis (Section~\ref{sec:covariates}) addresses several course-level factors; the finding that the association was carried mainly by differences between lecturers is consistent with this possibility, and the present data cannot distinguish legibility itself from such stable instructor-level practices.

    Of the three image-based indicators, only Readability consistently predicted student perceptions of clarity and organization. It reflects core aspects of visual accessibility: clarity of symbols, sharpness of writing, spacing, and overall legibility. Beyond its perceptual role, legible writing may also reflect the instructor's orientation toward the audience, that is, the extent to which the board is written for the students in the room rather than serving as a personal working space. This interpretation is consistent with the significant, if modest, association between Readability and attentiveness to students' questions ($r = .316$, $p < .05$), alongside its stronger associations with Clarity and Organization. Readability is also the dimension that can be judged most directly from a single static image, since whether symbols are legible does not depend on how the board developed during the lecture.

    Order, by contrast, did not show a meaningful association with any survey dimension (all $|r| < .08$), despite its positive correlation with Readability ($r = .336$, $p < .05$). Two complementary explanations may account for this pattern. The first is methodological. Order was the dimension on which human raters agreed least, and on which Claude's alignment with human judgment was weakest (31\% of the human ceiling; Table~\ref{tab:human_validation}). Moreover, interpreting spatial organization depends on the chronological development of the lesson, which a static snapshot does not convey: whereas text has a built-in reading order, the semantic order of board regions must be inferred. The second explanation is substantive: the relationship between board organization and student perception may not be monotonic. A disorganized board is likely to hinder students, yet a highly uniform, pre-planned layout may indicate a delivery that unfolds largely independently of the ongoing interaction with the class, for example when prepared notes are reproduced rather than adapted to students' questions and contributions. If both extremes are associated with lower ratings, a linear correlation would understate the relevance of Order. This interpretation remains speculative and should be examined in designs that test for non-linear effects.

    Visuals presents a different case. It was the most consistently measured dimension in this study: agreement among human raters was highest for Visuals ($\rho = .619$), and both AI models matched or exceeded this level in their agreement with the human mean (Table~\ref{tab:human_validation}). Yet Visuals was largely unrelated to students' ratings of clarity and organization. A plausible account is that the presence of diagrams, sketches, and color coding is determined primarily by the subject matter, as some topics are inherently diagrammatic while others rely almost entirely on symbolic derivation, rather than by how the instructor presents the material. Visuals thus appears to reflect a property of the content more than of the instruction, which would also explain why it is the least subjective dimension to rate. This pattern indicates that reliability of measurement does not in itself imply pedagogical relevance: the dimension on which human and automated raters converge most readily is the one least informative about student-perceived clarity. The modest association between Visuals and Interest ($r = .278$, $p < .05$) is consistent with this view, as topics that lend themselves to visual representation may also be experienced as more engaging.

    A related point concerns the selective nature of the image-based \textit{Readability} metric. Rather than acting as a broad proxy for teaching quality, \textit{Readability} exhibits its strongest associations with survey items that directly reflect visual and structural clarity in instruction, such as \textit{Clarity} ($r = .434, p < .01$) and \textit{Organization} ($r = .455, p < .01$). In contrast, it shows weaker associations with items reflecting affective or interpersonal aspects of instruction, such as the interest generated by the lecture ($r = .226$, n.s.) and, to a lesser extent, attentiveness to students' questions ($r = .316$). This pattern indicates that the metric captures the specific visual-organizational dimension it was designed to assess. From a construct perspective, such selectivity is a strength, enabling clear interpretability rather than collapsing into an undifferentiated signal of teaching effectiveness.
    
    This contribution is particularly meaningful given the internal structure of the student surveys themselves. Although all survey items were significantly intercorrelated (ranging from $r = .388$ to $r = .924$, all $p < .01$), prior work has often taken this coherence as evidence that a single global item may suffice for summative evaluation. Indeed, Cashin and Downey~\cite{cashin1992using}, analyzing ratings from 17,183 classes at 105 institutions, found that a single global item accounted for more than 50 percent of the variance in a weighted composite of student ratings, whereas student, class, and method items generally added much less. Global ratings thus summarize students’ overall judgment efficiently, but by design they do not indicate which specific aspect of instruction underlies that judgment.
    
    The present results speak to this limitation. Even within a highly homogeneous rating instrument, \textit{Readability} identifies differentiated variation linked to visual clarity and organization. Two instructors could receive highly similar global ratings while differing substantially in how clearly they structure information on the board. By uncovering variance that global scores tend to obscure, the \textit{Readability} metric demonstrates how AI-based visual analysis can surface instructional differences that matter for how information is processed and understood.

    The present findings provide initial evidence that visual assessment of lecture boards can serve as a practical and pedagogically meaningful source of information about how clearly instructional content is visually presented. Rather than attempting to capture the full complexity of instruction, image-based analysis offers a focused indicator of how clearly information is visually communicated to students. This complements existing evaluation approaches by directly targeting a dimension of instructional practice that students regularly encounter but that has been difficult to assess systematically at scale.
    
    Three potential implications follow from this result. First, lecture boards represent an accessible and low-cost substrate for continuous monitoring of key visual features in STEM instruction, enabling scalable evaluation without requiring manual annotation or intrusive observation. Second, by capturing visual clarity independently of content, the \textit{Readability} metric offers a complementary perspective to student surveys, helping institutions separate perceptual aspects of instruction from broader affective or interpersonal judgments. Third, with further refinement, such metrics could support formative pedagogical reflection: once the reliability of the measurement is improved, instructors could receive targeted feedback on aspects of their board writing that are associated with perceived clarity, prompting simple, actionable improvements.
        
    Together, these implications suggest that automated visual analysis should not be viewed as a replacement for existing teaching evaluation systems but as an additional layer that highlights specific and interpretable aspects of instructional practice. As AI capabilities continue to expand, integrating such measures with other sources of pedagogical information may enable more nuanced and equitable evaluation frameworks that reflect how students actually experience instruction in the classroom.

    The present study should be viewed as a proof of concept: its aim was to examine the feasibility of treating lecture board writing as an automatically measurable artifact, and whether such a measure carries a signal that aligns with student perceptions. The findings support this feasibility, but several limitations should be acknowledged when interpreting the results.
        
    First, the use of board images necessarily focuses on the visual-spatial dimension of instruction: the analysis is centered on visual clarity, organization, and legibility. Consequently, it does not fully capture other critical aspects of teaching, such as interpersonal instructor–student interaction, responsiveness to questions, and classroom climate. Image-based evaluation should therefore be understood as a decision-support tool that highlights structural strengths and supports pedagogical reflection, rather than a standalone metric of teaching effectiveness.

    Second, the measurement of the dimensions themselves contains considerable noise. The prompt produced highly stable scores across repeated runs (Sections~\ref{sec:stab_single}--\ref{sec:stab_within}), but stability across runs does not guarantee agreement with human judgment. Agreement among human raters on Readability, the dimension on which the main findings rest, was moderate (mean leave-one-out $\rho = .446$, with one rater at .198), and Claude's alignment with the human mean reached 64\% of that ceiling; for Order, alignment was considerably lower (31\%). This suggests that the constructs, and not only the scoring method, are defined loosely enough to be interpreted differently by different observers. To the extent that this noise is random, it would attenuate rather than inflate the observed associations with student evaluations. Nonetheless, future studies will need to develop prompts with more explicit operational definitions of each dimension, and to establish higher human--human and human--AI agreement before such measures are applied more broadly.
    
    Third, student evaluations of teaching (SETs) serve here as the reference measure against which the AI-based scores were compared. SETs remain the only systematically collected indicator of student perception available at scale, and their use in this study is pragmatic: they capture how students experienced the course, which is precisely the construct the visual metric is intended to approximate. Because SET scores are collected once per course and depend on voluntary participation, we restricted analyses to courses with adequate survey response rates (median 44\%, range 16--81\%) in order to improve the representativeness of the reference. While the results demonstrate meaningful convergence with student perceptions, future research incorporating direct measures of student learning would further illuminate the instructional significance of these visual metrics.
    
    Fourth, despite leveraging a large number of raw observations (including 2,645 images and over 130,000 AI inference samples) the statistical comparison to survey outcomes operates at the course level, as each course contributes a single aggregated SET score. This results in a sample of $N=56$ course instances taught by 18 lecturers. While adequate for an initial feasibility study and supported by extensive robustness analyses, this sample is drawn from a single department at a single institution. Extending the analysis to additional courses, disciplines, and institutions will be essential for assessing generalizability and for modeling instructor-level variation across pedagogical styles.
        
    Finally, applying AI to analyze real instructional artifacts introduces ethical and governance considerations, including data ownership, instructor consent, and the potential for reductive or punitive use of automated metrics. These issues extend beyond the scope of the current manuscript but are crucial for responsible future deployment. Institutional policies must ensure transparency, instructor agency, and safeguards against misuse of AI-driven evaluative data.

    Beyond these limitations, a natural next step is a controlled experiment that tests the effect of board writing on student understanding directly. The same lecture content could be presented on boards that vary systematically in readability and in organization, for example in a design crossing legible and less legible writing with organized and disorganized layouts, while students' comprehension and recall are measured alongside their perceived clarity. Such a design would allow causal inference, which the present correlational data do not support, and would separate the contributions of readability and organization. Including intermediate levels of organization would also make it possible to test whether the relationship between board order and student outcomes is non-monotonic, as suggested above.
    
    Taken together, these limitations indicate that AI-based visual metrics should be adopted cautiously and contextually, as part of a broader multi-source evaluation ecosystem that respects both the complexity of teaching and the professional autonomy of instructors.

    In conclusion, this study provides proof-of-concept evidence that AI-driven analysis of lecture-board images can serve as a scalable, artifact-based indicator of visual instructional clarity in quantitative STEM education. Across 56 university courses, the AI-derived board Readability score was associated with student-rated clarity and organization, and these associations remained stable under thin-slice sampling, indicating that estimates can be obtained from a small number of images per lecture.

    The central finding, that the readability of board writing is reflected in how students perceive the clarity of instruction, suggests that the lecture board can be treated as a quantifiable instructional artifact and offers a starting point for future research. The methods introduced here may provide a low-cost complement to existing evaluation practices. Future work should refine the measurement of each dimension to achieve higher agreement with human judgment, test the effect of board writing on learning in controlled settings, extend the framework across institutions and disciplines, and explore governance models that ensure transparent, ethical, and formative use of AI-assisted teaching analytics.    
    
    \section{Methods}
    \label{sec:Methodology}
    
    Recognizing that the lecture board constitutes a central pedagogical tool in the instruction of physics and mathematics, in this study, we developed an AI-based automated tool for the evaluation of academic lecture boards. We examined 56 academic courses offered by Bar-Ilan University between 2017 and 2023. To capture variation across course types, years, and lecturers, we selected both physics content courses and mathematics courses taught within the physics program, across various years, featuring different lecturers and topics, and including both mandatory and elective courses. Additionally, the diversity of camera angles and lecture halls across different courses supports the generality of the findings. Importantly, all selected sessions possessed uniform video recording quality, and the viewpoint within each individual course remained largely consistent. Furthermore, because SET scores depend on voluntary participation and are
    therefore sensitive to response rate, only courses with adequate student response rates (median 44\%, range 16--81\%) were included; see the full table of courses in Supplementary Table S1. This criterion ensured that the SET scores used as a reference measure more accurately reflected the perceptions of the enrolled cohort.
    Notably, the lecture board images were extracted retrospectively from archival recordings produced as part of routine instructional activity. At the time of teaching, instructors were not aware that their board writing would later be analyzed for research purposes, minimizing observer effects and reducing the risk of behavior modification in response to evaluation. Instructors provided written informed consent for the research use of their archived recordings prior to any data extraction or analysis, and all analyses were conducted on anonymized course-level data.
    
    We classified the courses according to academic progression: the first-year courses were designated as the study year 1, second-year courses as the study year 2, and the third-year and graduate courses as the study year 3. See Table~\ref{tab:stats} for further details on course classification.
    
    To utilize the lecture board data effectively, frames of the lecture board were extracted from the archived lecture recordings at 20, 40 and 60 minutes from the beginning of the lecture (see Subsection~\ref{subsec: slicing}). Each image underwent manual inspection to ensure consistent quality and to verify the visibility of the board. In instances where the board was obscured by presentation slides or an experimental setup, an alternative image was selected from a time window of $\pm 5$ minutes relative to the original timestamp; if no suitable image was found, the data point was discarded.
    
    Image analysis was performed using a standardized Python automation script designed to maintain a consistent prompt structure. Each image was processed by the AI model 50 times, and the average of these iterations was recorded as the final result to ensure statistical robustness. For the AI model, we utilized Anthropic’s Claude 3.5 Sonnet (version 20241022) via API. This specific model was selected as one of the leading solutions capable of processing image inputs through an API. This automated workflow was implemented to eliminate human manipulation or bias during the data evaluation phase.

    \begin{table}[h!]
        \caption{Course Statistics and Data Summary}
        \label{tab:stats}
        \begin{tabular}{lc}
            \toprule
            \textbf{Classification} & \textbf{Count} \\
            \midrule
            Study Year One Courses & 24 \\
            Study Year Two Courses & 20 \\
            Study Year Three Courses & 12 \\
            \midrule
            Physics Content Courses & 40 \\
            Mathematics for Physicists Courses & 16 \\
            \midrule
            Total Unique Courses & 23 \\
            Total Different Lecturers & 18 \\
            \midrule
            \textbf{Total Courses (Entries)} & \textbf{56} \\
            \textbf{Total Lectures} & \textbf{910} \\
            \textbf{Total Pictures} & \textbf{2,645} \\
            \bottomrule
        \end{tabular}
    \end{table}

    \subsection{The prompt}
    \label{prompt}
    
    In the process of developing a reliable prompt for evaluating lecture boards, our goal was to create a tool that is both stable and reliable and at the same time capable of highlighting the most pedagogically significant features of a board. Initially, a general prompt was used without specific attributes and two representative images were selected from two courses. Those courses had received substantially different teaching evaluation scores in order to capture a wide range of board quality. For each image, the model was instructed to act as an expert in evaluating teaching quality and to provide both a score of 1 to 10 for the board and a short explanation of its score, while focusing only on the visualization of the board as follows.
    
    \textit{``Act as an expert in evaluating the quality of teaching in academia.}
    
    \textit{The image shows a lecture board. Ignore the content and provide two things:}
    
    \textit{1.A numerical score from 1 to 10 for the board's visualization quality.(give the number in this way: n, and not in this way:n/10(also give the number as the first in the string)}
    
    \textit{2. A short explanation about the board's visualization quality.''}
    
    The instruction to ignore the content was not intended to dismiss the pedagogical importance of what was written on the board. Rather, it served as a methodological control designed to isolate the visual component of board writing. By directing the model not to evaluate the mathematical or conceptual correctness of the material, the prompt focused exclusively on properties such as handwriting legibility, spatial organization, and visual clarity - separating the visual accessibility of the board from the semantic quality of the lecture content.
    
    Each image was run 500 times to assess stability, yielding a mean of 7.348, a standard deviation of 0.476, and a coefficient of variation (CV) of 6.489\% for the first image, and a mean of 4.16, a standard deviation of 0.75, and a CV of 18.045\% for the second image. The low CV values indicate that the model produced consistent numerical evaluations.
    
    The qualitative explanations were used to define the parameters evaluated by the Large Language Model (LLM) acting as a pedagogical expert in visual analysis. Initially, the explanations were processed into a word cloud to identify the most prevalent terms. Supplementary Fig.~S1 displays these high-frequency words and their subsequent classification into three distinct categories. This categorization process was performed by the authors. The categories were: Readability and Clarity, Organization and Layout, and Visual Elements and Illustration each on a scale from 1 to 10, and to provide only the three numerical scores without additional explanation.
    
    \textit{``Act as an expert in evaluating the quality of teaching in academia.}
    
    \textit{The image shows a lecture board. Ignore the content and rate the following three categories from 1 to 10:}
    
    \textit{a. Readability and Clarity.}
    
    \textit{b. Organization and Layout.}
    
    \textit{c. Visual Elements and Illustration}
    
    \textit{.where 1 is the lowest score. Provide only 3 numbers, one for each category, without any explanation.''}
    
    For simplicity, the three categories will henceforth be referred to as \textit{Readability}, \textit{Order}, and \textit{Visuals}.

    Although the prompt positions the model as an expert in teaching evaluation, its output is constrained to three visual dimensions; the framing serves to anchor the evaluative perspective, not to elicit a judgment of teaching quality as such.
    
    \subsection{Stability and Reliability of the Prompt in a Single Image}
    \label{sec:stab_single}

    After systematically refining and standardizing the prompt parameters and formulating three evaluation questions with fixed wording, we examined whether the finalized prompt exhibits statistical stability and consistency across repeated runs for the same image. To test this, we selected an image that was considered high-quality and ran the prompt 5,000 times.
    Evaluation scores (on a scale of 1-10) were distributed as follows: for the \textit{Readability} parameter, a score of 9 was obtained in 4,324 cases and 8 in 676 cases; for \textit{Order}, 9 in 664 cases, 8 in 4,325 cases and 7 in 11 cases; and for \textit{Visuals}, 9 in 66 cases, 8 in 3,041 cases and 7 in 1,893 cases.
    For all three parameters (\textit{Readability}, \textit{Order}, and \textit{Visuals}), the results demonstrate a high degree of stability, with the vast majority of scores concentrated around the upper end of the scale. This indicates that the prompt yields highly consistent evaluations across repeated runs.

    \subsection{Consistency Analysis of AI Scoring Across Courses}
    
    After confirming that the prompt produces consistent and stable results for a single specific image, we evaluate whether it remains reliable across different courses and instructors, reflecting varying teaching styles. This analysis is crucial to ensure that the prompt can be generalized beyond individual cases and does not depend on a specific board or lecturer's style.

    To perform this assessment, a random representative sample consisting of 32 courses was selected and a single time point, identical for all courses, was randomly chosen from which images were extracted.
    
    Each selected image was processed by the AI model 150 times to obtain sufficient repetitions for a robust evaluation of stability. The results were quantified using the coefficient of variation (CV) for each evaluation parameter. The mean CV for \textit{Readability} was 4.85\%, with a minimum of 0\% and a maximum of 12.81\%. For \textit{Visuals}, the mean CV was 8.64\%, with a minimum of 1.36\% and a maximum of 16.26\%. For \textit{Order}, the mean CV was 5.31\%, with a minimum of 0\% and a maximum of 12.01\%.
    
    These results indicate that the prompt maintains a high degree of consistency even when applied to the lecture boards of different instructors and courses, that is, in a range of teaching styles. Notably, \textit{Readability} and \textit{Order} showed particularly low variability, while \textit{Visuals} exhibited slightly higher but still acceptable variation, suggesting that the prompt is robust in evaluating core aspects of board quality regardless of stylistic differences. This finding supports the generalization of the prompt as a consistent tool for characterizing the visual properties of the lecture board in diverse educational settings.
    
    \subsection{Stability and Reliability of the Prompt Within a Single Course}
    \label{sec:stab_within}
    
    To evaluate the stability of the prompt within each individual course, for each lecture in each of the 56 courses, three images were selected according to the slicing assumption (see Section~\ref{subsec: slicing}), and each image was processed by the AI model in 50 independent runs.
    We used the coefficient of variation (CV) and standard deviation (SD) values to quantitatively assess the consistency of the scores within each course under the prompt.
    
    For the \textit{Readability} parameter, the mean SD was 0.64 (minimum 0.23, maximum 1.39) and the mean CV was 8.74\% (minimum 2.9\%, maximum 21.25\%). For \textit{Order}, the mean SD was 0.58 (minimum 0.20, maximum 1.16) and the mean CV was 8.20\% (minimum 2.79\%, maximum 16.79\%). For \textit{Visuals}, the mean SD was 0.98 (minimum 0.51, maximum 1.47) and the mean CV was 16.56\% (minimum 7.78\%, maximum 28.58\%). The general averages in all three parameters were a mean SD of 0.73 (minimum 0.20, maximum 1.47) and a mean CV of 11.16\% (minimum 2.79\%, maximum 28.58\%), indicating that, in general, the scores within each course remained consistent across lectures, with \textit{Visuals} showing slightly higher variability compared to \textit{Readability} and \textit{Order}.
     Those outcomes provide a clear view of how consistent the prompt's evaluations were within individual courses and support the use of course-level averages.

    \subsection{Sample Size}
    Large language models (LLMs) are known to produce non-identical responses even when given the same prompt independently multiple times. Therefore, we sought to determine how many runs per image are required for the mean of all runs to be meaningful, stable, and reliable.
    To investigate this, we conducted the following analysis: we took the results of a single image that had been run 5000 times and randomly selected subsets of increasing size. For each subset, we calculated the standard deviation (SD) for each evaluation parameter in the prompt, see Supplementary Fig.~S2. The graph shows that for all three parameters, the SD decreases approximately in proportion to $\frac{1}{\sqrt{N}}$, where N is the subset size, as accepted according to the Central Limit Theorem (CLT). Moreover, starting at around 50 runs, the slope of the curve becomes very small and no longer significant.
    These results indicate that even a relatively small number of runs, around 50, is sufficient to obtain stable and reliable mean scores across all three parameters. This finding informed our choice of 50 runs per image for the stability analysis within the course.

    \subsection{Survey Data Collection}\label{subsec: data collection}
    
    In this study, the AI-based board scores were compared with institutional teaching evaluation surveys conducted at Bar-Ilan University from 2017 to 2023. Although the wording and number of questions changed slightly over the years, we selected the central items that consistently appeared across surveys in order to ensure comparability. In addition, we computed a weighted average score based on all available items in each survey, providing a single summary indicator of students' overall evaluation of the course.
    
    The six variables used in our analyses are listed below with their original wording as presented to students, together with the abbreviated labels by which we refer to them later in the manuscript:
    
    \begin{itemize}
        \item \textbf{Clarity} – ``The lecturer taught in a clear manner.''
        \item \textbf{Interest} – ``The lecturer taught in an interesting manner.''
        \item \textbf{Organization} – ``The course was organized and structured.''
        \item \textbf{Questions} – ``The lecturer was attentive to students’ questions during and after class.''
        \item \textbf{Satisfaction} – ``Overall, I am satisfied with the teaching in this course.''
        \item \textbf{Average} – Weighted mean of all survey items.
    \end{itemize}
    
    In the following sections, these survey-based variables are referred to by their abbreviated labels (e.g., \textit{Clarity}, \textit{Interest}).

    \subsection{Rationale for Sampling Time Points and Thin Slicing}\label{subsec: slicing}
    
    The three time points (20, 40, and 60 minutes) were selected to capture different stages of a typical lecture – early exposition, mid-lecture development, and later consolidation – while minimizing the number of samples required. Although a single image may appear to represent only a fleeting moment, in practice it condenses the instructional activity that accumulated on the lecture board up to that point in time, thereby offering a meaningful snapshot of the lecture’s evolving structure, organization, and clarity.
    
    This design is closely connected to the principle of thin slicing \citep{ambady1993half}, which addresses the challenge of drawing reliable conclusions from minimal data. In this study, we test whether student-perceived clarity can be predicted from a reduced set of lecture board images, and to what extent such predictions depend on the stability of instructors' visual presentation over time. Put differently, the question of whether accurate assessments can be derived from fewer images is essentially equivalent to asking how stable instructors are in their teaching behavior – whether they consistently maintain handwriting clarity, board organization, and overall legibility, or whether large variations across a lecture require more extensive sampling. Importantly, we also examine whether only part of the images – or even a single snapshot – may suffice for reliable prediction, rather than requiring all three time points. Demonstrating the feasibility of such thin-slice predictions would highlight that even limited visual evidence carries a signal that aligns with student perceptions, offering a scalable and efficient pathway for AI-assisted analysis of instructional artifacts.

    \subsection{Statistical analysis}

    Associations between AI-derived and survey measures were assessed with two-tailed Pearson correlations; Bonferroni correction was applied across the 18 AI--survey correlations. Covariate analyses used hierarchical linear regression, with standard errors clustered by lecturer as a robustness check. Agreement between human raters and AI scores was assessed with Spearman correlations.

    \subsection{Ethics}
    \label{sec:ethics}

    The study was approved by the Human Subjects Institutional Review Board of Bar-Ilan University (approval no. 310825718) and was performed in accordance with relevant guidelines and regulations. All instructors whose lectures were included provided written informed consent for the research use of their archived lecture recordings prior to any data extraction or analysis. Because the recordings were made as part of routine instruction, before the study was conceived, instructors were unaware at the time of teaching that their board writing would later be analyzed. Student evaluation data were obtained from institutional records as course-level aggregates only: the mean score of each survey item, the response rate, the number of enrolled students and the mean course grade. No names, student identifiers, demographic information or individual responses were accessed at any stage. The need for informed consent from students was waived by the Human Subjects Institutional Review Board of Bar-Ilan University. No identifying information about students or instructors is reported.

    \subsection{Code availability}

    The code used in this study is available from the corresponding author on reasonable request.

    \section*{Data availability}

    The anonymized data supporting the findings of this study are openly available in Zenodo at https://doi.org/10.5281/zenodo.23452528. The board images extracted from the lecture recordings are not publicly available to protect the privacy of instructors and students. Requests for data should be addressed to the corresponding author, Yossi Ben Zion (benzioy@biu.ac.il).

    \bibliography{ref}

    \section*{Acknowledgements}

    We thank the lecturers who agreed to the research use of their recorded lectures, and Prof. Gerd Kortemeyer (ETH Zurich) for helpful comments on this work. This research received no external funding.

    \section*{Author contributions}

    N.H.: data curation, formal analysis, investigation, resources, software, validation, visualization, writing -- original draft, writing -- review \& editing. R.S.: formal analysis, investigation, resources, validation, writing -- original draft, writing -- review \& editing. Y.B.Z.: conceptualization, formal analysis, methodology, supervision, writing -- original draft, writing -- review \& editing. All authors reviewed and approved the final manuscript.

    \section*{Additional information}

    \textbf{Competing interests:} The authors declare no competing interests.

    \textbf{ORCID iDs:} Nachshon Hirschenberger \url{https://orcid.org/0009-0006-3637-6415}; Roni Segal \url{https://orcid.org/0009-0002-8606-3442}; Yossi Ben Zion \url{https://orcid.org/0000-0002-9337-9392}.

    % ---- Supplementary Information (merged from supplementary.tex for arXiv) ----
    \clearpage
    \setcounter{table}{0}
    \setcounter{figure}{0}
    \renewcommand{\thetable}{S\arabic{table}}
    \renewcommand{\thefigure}{S\arabic{figure}}
    \renewcommand{\theHtable}{S\arabic{table}}
    \renewcommand{\theHfigure}{S\arabic{figure}}
    \renewcommand{\tablename}{Supplementary Table}
    \renewcommand{\figurename}{Supplementary Fig.}

\begin{center}
{\Large\bfseries Supplementary Information}\\[6pt]
{\large Lecture Board Readability as an AI-Measurable Predictor of Perceived Teaching Clarity}\\[6pt]
Nachshon Hirschenberger, Roni Segal, Yossi Ben Zion\\
Department of Physics, Bar-Ilan University, Ramat Gan, Israel
\end{center}

    \begin{table}[h!]
    \centering
    \caption{\label{tab:S_courses}Course details and survey response rates.}
    \footnotesize
    \renewcommand{\arraystretch}{0.85}
    \begin{tabular}{ccccc}
    \toprule
    Course Code & Year & Year of Study & Lecturer Code & Response Rate (\%) \\
    \midrule
    A1 & 2017 & 2 & L16 & 77.33 \\
    A2 & 2018 & 2 & L16 & 57.5 \\
    A3 & 2022 & 2 & L1 & 39.56 \\
    A4 & 2023 & 2 & L1 & 21.05 \\
    B & 2023 & 3 & L12 & 47.95 \\
    C & 2017 & 2 & L6 & 78.05 \\
    D1 & 2019 & 1 & L7 & 42.59 \\
    D2 & 2023 & 1 & L10 & 22.8 \\
    D3 & 2021 & 1 & L10 & 31.03 \\
    D4 & 2022 & 1 & L10 & 23.8 \\
    E1 & 2017 & 1 & L1 & 48.39 \\
    E2 & 2019 & 1 & L1 & 43.59 \\
    E3 & 2023 & 1 & L12 & 43.01 \\
    E4 & 2021 & 1 & L12 & 29.83 \\
    E5 & 2022 & 1 & L13 & 28.88 \\
    E6 & 2022 & 1 & L12 & 25.6 \\
    F & 2019 & 3 & L4 & 58.33 \\
    G1 & 2017 & 1 & L12 & 51.35 \\
    G2 & 2022 & 1 & L14 & 27.77 \\
    G3 & 2023 & 1 & L14 & 27.63 \\
    H2 & 2023 & 1 & L5 & 41.82 \\
    I1 & 2017 & 1 & L8 & 81.25 \\
    I2 & 2023 & 1 & L3 & 49.12 \\
    I3 & 2020 & 1 & L3 & 50.87 \\
    I4 & 2020 & 1 & L6 & 39.6 \\
    I5 & 2022 & 1 & L3 & 37.2 \\
    J1 & 2023 & 2 & L1 & 51.9 \\
    J2 & 2020 & 2 & L1 & 59.49 \\
    J3 & 2022 & 2 & L1 & 43.75 \\
    K1 & 2022 & 3 & L10 & 34.07 \\
    K2 & 2020 & 3 & L10 & 44.44 \\
    K3 & 2023 & 3 & L10 & 52.42 \\
    M & 2018 & 3 & L15 & 51.11 \\
    N1 & 2017 & 2 & L1 & 80.6 \\
    N2 & 2018 & 2 & L1 & 74.7 \\
    N3 & 2022 & 2 & L9 & 39.84 \\
    N4 & 2023 & 2 & L9 & 51.69 \\
    O & 2019 & 2 & L5 & 47.37 \\
    P1 & 2017 & 2 & L2 & 58.57 \\
    P2 & 2018 & 2 & L2 & 58.93 \\
    P3 & 2019 & 2 & L2 & 40 \\
    P4 & 2021 & 2 & L2 & 19.62 \\
    P5 & 2022 & 2 & L2 & 30.83 \\
    P6 & 2023 & 2 & L2 & 20.4 \\
    Q & 2023 & 3 & L2 & 43.75 \\
    R1 & 2022 & 3 & L11 & 29.7 \\
    R2 & 2017 & 3 & L12 & 65.15 \\
    S & 2017 & 2 & L2 & 66.1 \\
    T1 & 2021 & 3 & L1 & 50 \\
    T2 & 2023 & 3 & L1 & 46.15 \\
    U & 2022 & 1 & L17 & 27.02 \\
    V1 & 2021 & 1 & L17 & 30.9 \\
    V2 & 2022 & 1 & L17 & 25 \\
    V3 & 2023 & 1 & L18 & 16 \\
    W & 2020 & 1 & L5 & 56.86 \\
    X & 2020 & 3 & L15 & 51.51 \\
    \bottomrule
    \end{tabular}

    \vspace{3pt}
    \parbox{0.75\textwidth}{\footnotesize Course names and instructors are anonymized; labels such as A1--A4 denote instances of the same course in different academic years. Year of study: 1--3 (3 = third-year and graduate courses).}
    \end{table}

\clearpage
\section*{Supplementary Figures}

    \begin{figure}[h!]
        \centering
        \includegraphics[width=0.7\linewidth]{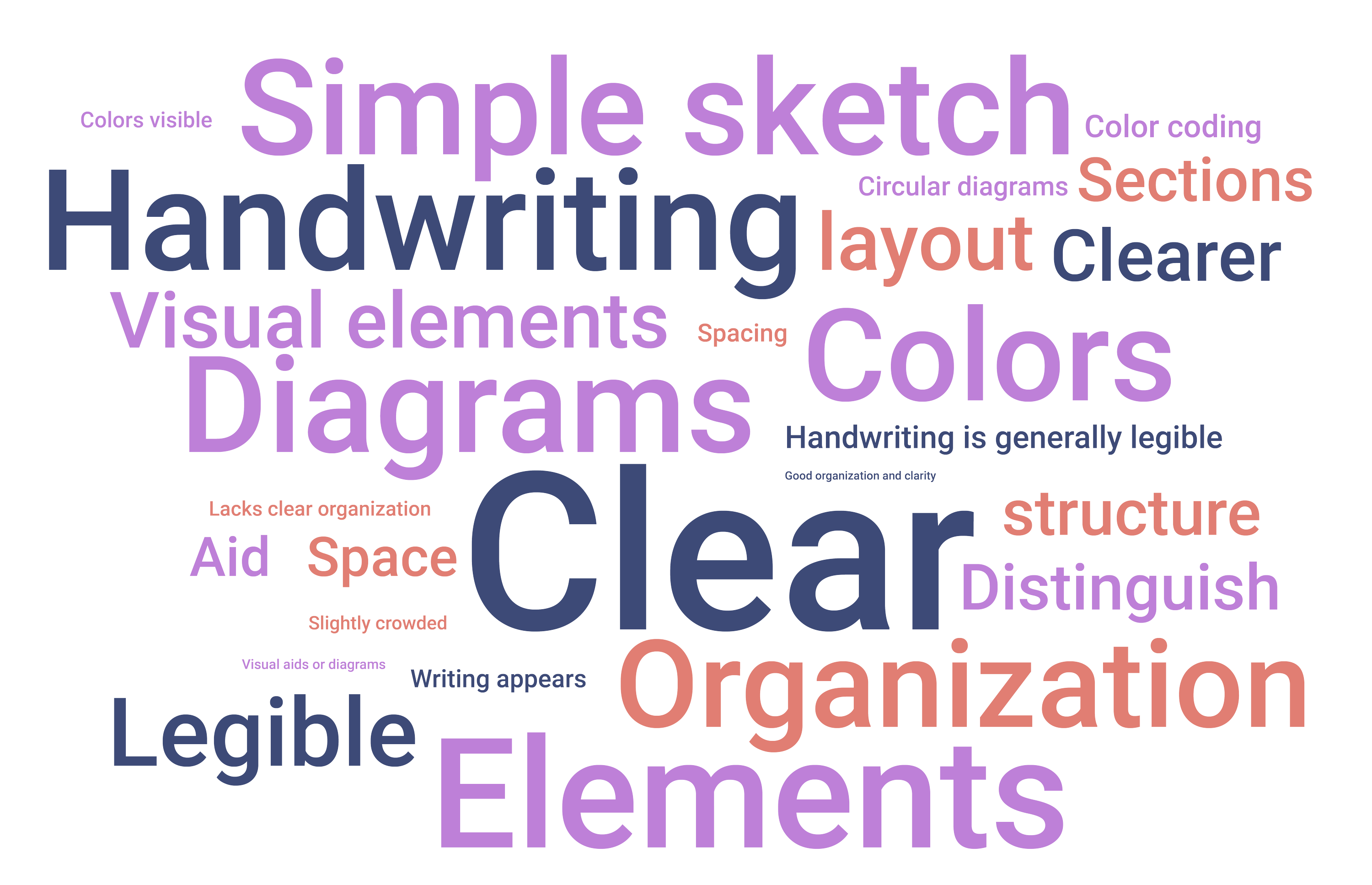}
        \caption{Word cloud visualization derived from 1,000 AI-generated explanations, illustrating the semantic extraction of the three evaluation parameters.}
        \label{fig:S_wordcloud}
    \end{figure}

    \begin{figure}[h!]
        \centering
        \includegraphics[width=\textwidth]{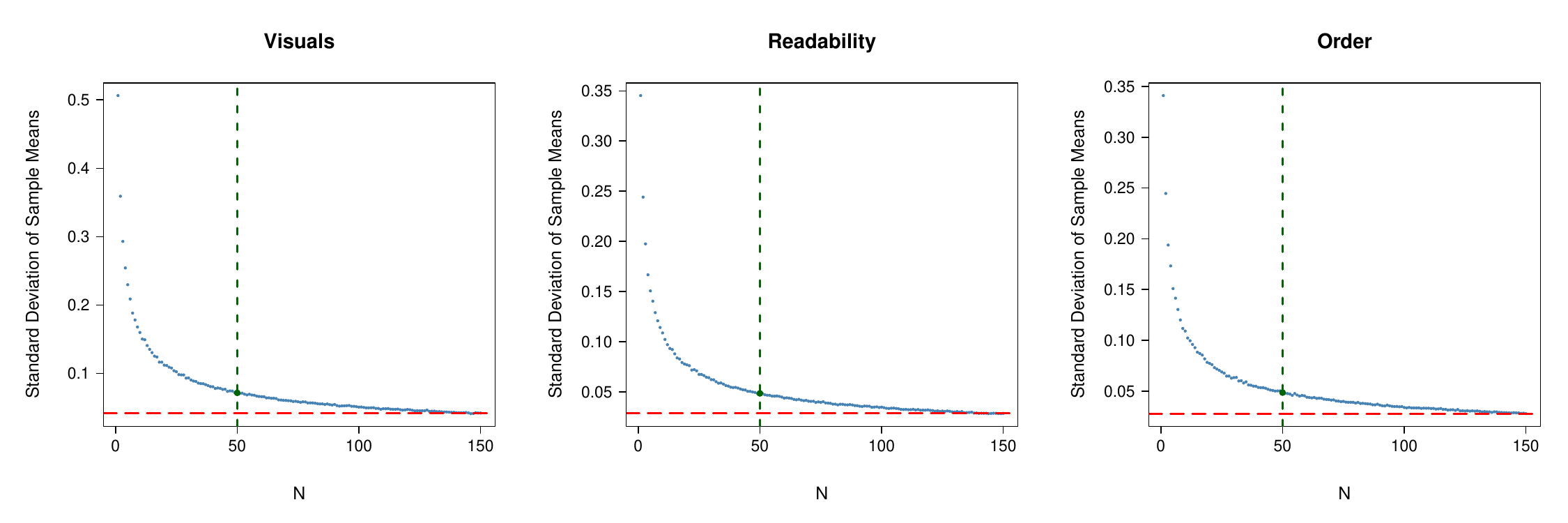}
        \caption{The graphs display the Standard Deviation of Sample Means as a function of subset size ($N$) for the three evaluation parameters: \textit{Visuals}, \textit{Readability}, and \textit{Order}. The standard deviation decreases approximately in proportion to $1/\sqrt{N}$, consistent with the Central Limit Theorem. The vertical green dashed line at $N=50$ marks the point where the curve stabilizes, indicating that 50 runs are sufficient to obtain reliable mean scores.}
        \label{fig:S_runs}
    \end{figure}

\clearpage
\section*{Supplementary Note 1: Exploratory multimodal analysis}

    This note reports a supplementary, exploratory analysis conducted on a subsample of courses. It lies outside the research questions of the present study, and its results are preliminary and require confirmation in future research.

    \subsection*{Data}
    The analysis was conducted on a subsample of 32 courses from the present study. Board-based scores were combined with transcript-based indicators from a previous study \cite{zion2025ai}, in which lecture audio was transcribed using Whisper large-v2 and each lecture was evaluated 50 times by GPT-4o Mini. Transcript-based scores were available for 30 of the 32 courses. Of the two transcript-based dimensions in that study, \textit{Clarity of Explanation} and \textit{Lecture Structure and Logical Flow}, we use the latter (henceforth \textit{Order-Text}); the two dimensions were highly collinear ($r = .97$) and showed similar associations with student evaluations.

    \subsection*{Results}
    Correlations are presented in Supplementary Table~\ref{tab:S_multimodal}. Both Readability-Image and Order-Text were significantly correlated with student-rated Clarity and Organization, and neither was significantly correlated with attentiveness to students' questions. The two measures were positively but not significantly correlated with each other ($r = .319$, $p = .085$), suggesting that they may capture partly distinct aspects of instruction. Their product (Order-Text $\times$ Readability-Image) was associated with Clarity ($r = .585$; $R^2 = .342$, $F(1,28) = 14.58$, $p < .001$). Because the product term is highly correlated with each of its components ($r = .792$ and $r = .831$), this association should not be interpreted as evidence of an interaction between the two channels.

    \begin{tableorg}[ht]% plain LaTeX table: sn-jnl's table wrapper (threeparttable) breaks \resizebox
    \centering
    \caption{\label{tab:S_multimodal}Correlations between survey-based measures and AI-derived measures from the present (image-based) and previous (transcript-based) studies, exploratory subsample (N=32; correlations involving Order-Text or the product term are based on the 30 courses with transcript data).}
    \resizebox{\textwidth}{!}{
    \begin{tabular}{l|l|cccccc|ccc}
    \hline
    & & \multicolumn{6}{c|}{Survey} & \multicolumn{3}{c}{AI} \\
    \midrule
    & & Clarity & Interest & Organization & Questions & Satisfaction & Average & Readability-Image & Order-Text & Order $\times$ Readability \\
    \midrule
    \multirow{6}{*}{\rotatebox{90}{Survey}}
    & Clarity       & 1       &        &        &        &        &        &        &        &        \\
    & Interest      & .750**  & 1      &        &        &        &        &        &        &        \\
    & Organization  & .811**  & .620** & 1      &        &        &        &        &        &        \\
    & Questions     & .613**  & .380*  & .443*  & 1      &        &        &        &        &        \\
    & Satisfaction  & .928**  & .834** & .793** & .676** & 1      &        &        &        &        \\
    & Average       & .918**  & .782** & .848** & .638** & .931** & 1      &        &        &        \\
    \hline
    \multirow{3}{*}{\rotatebox{90}{AI}}
    & Readability-Image   & .520**  & .287   & .449** & .311   & .438*  & .424*  & 1      &        &        \\
    & Order-Text          & .431*   & .402*  & .413*  & .155   & .435*  & .405*  & .319   & 1      &        \\
    & Order $\times$ Readability  & .585**  & .415*  & .536** & .282   & .530** & .521** & .792** & .831** & 1      \\
    \hline
    \end{tabular}
    }
    \begin{flushleft}
    \textit{Note:} Correlations are Pearson coefficients, computed with pairwise deletion. Statistically significant values are marked with $^{*}p<.05$, $^{**}p<.01$ (two-tailed).
    \end{flushleft}
    \end{tableorg}

    \subsection*{Interpretation}
    These preliminary findings suggest that board-based and transcript-based measures may provide partly complementary information about students' perceptions of instructional clarity. Given the small subsample and the exploratory nature of the analysis, this possibility should be examined in larger multimodal studies before conclusions are drawn. The board-based measure developed in the present study does not depend on such integration and can be applied where transcripts are unavailable.

    \end{document}